\documentclass[twocolumn,showpacs,amsmath,amssymb]{revtex4}
\usepackage{graphicx,epsfig}
\usepackage{dcolumn}
\usepackage{bm}
\newcommand{\beq}{\begin{equation}}
\newcommand{\eeq}{\end{equation}}
\newcommand{\bseq}{\begin{subequations}}
\newcommand{\eseq}{\end{subequations}}
\newcommand{\beqa}{\begin{eqnarray}}
\newcommand{\eeqa}{\end{eqnarray}}
\newcommand{\beqas}{\begin{eqnarray*}}
\newcommand{\eeqas}{\end{eqnarray*}}

\begin{document}

\title{Symmetry, winding number and topological charge of vortex solitons
in discrete-symmetry media}

\author{Miguel-{\' A}ngel Garc\'{i}a-March$^1$, Albert Ferrando$^2$, Mario Zacar\'es$^1$,  Sarira
  Sahu$^3$, and Daniel E. Ceballos-Herrera$^{1,4}$}

\affiliation{
$^1$Institut Universitari de Matem\`atica Pura i Aplicada - IUMPA, Universitat Polit\`ecnica de Val\`encia,
Cam\'{\i} de Vera, s/n. E-46022 Val\`encia, Spain.\\
$^2$Departament d'{\'O}ptica, Universitat de Val\`{e}ncia, Dr. Moliner,
50. E-46100 Burjassot (Val\`{e}ncia ), Spain\\
$^3$Instituto de Ciencias Nucleares, Universidad Nacional Aut\'onoma de
M\'exico,  
Circuito Exterior, C.U., A. Postal 70-543, 04510 M\'exico DF, M\'exico.\\
$^4$Centro de Investigaciones en \'Optica A.C, Le\'on, Gto. 37170, M\'exico.
}

\begin{abstract}
We determine the functional behavior near the discrete rotational symmetry axis of discrete vortices of the nonlinear Schr\"odinger equation. We show that these solutions present a central phase singularity whose charge is restricted by symmetry arguments. Consequently, we demonstrate that the existence of high-charged discrete vortices is related to the presence of other off-axis phase singularities, whose positions and charges are also restricted by symmetry arguments.  To illustrate our theoretical results, we offer two numerical examples of high-charged discrete vortices in photonic crystal fibers showing hexagonal discrete rotational invariance.

\end{abstract}

\maketitle

\section{Introduction}

Complex scalar solutions of wave equations can present dislocations similar to those found in crystals~\cite{Nye74}. The essential mathematical property of these complex scalar functions  in the point or line where a dislocation is localized is that its phase is increased or decreased in a multiple of $2\pi$ along a closed curve around it~\cite{Nye74}. In these points or lines, also known as phase singularities, the amplitude of the function vanishes and its phase is undetermined.  They play an important role in many branches of science,  such as solid-state physics~\cite{Kos73}, Bose-Einstein condensates (BEC)~\cite{Mad00}, superfluidity~\cite{Sal87}, superconductivity~\cite{Bla94}, cosmology~\cite{Bar07}, molecular dynamics~\cite{Ald70}, nonlinear optics~\cite{DesyatnikovPO2005}, etc. In the latter case, the study of such singularities is often enclosed in a separated branch called nonlinear singular optics~\cite{SoskinPAO1998}. 

An optical vortex is a complex scalar solution of a wave equation defined in a two-dimensional domain characterised by the presence of a phase singularity~\cite{DesyatnikovPO2005}. If the vortex has a circular symmetry there is only a  singularity located at the rotational axis. 
Optical vortices with discrete symmetry, or discrete vortices (DV's), have been theoretically predicted to exist in inhomogeneous periodic media as solitonic solutions of a nonlinear wave equation, such as the  nonlinear Schr\"odinger equation (NLSE). These periodic media include optically induced lattices~\cite{Yan03}, photonic crystal fibers~\cite{Fer04} or Bessel lattices~\cite{KartashovPRL2005}. They have been experimentally observed in the former medium~\cite{Nes04,Fle04}. On the other hand,  DV's have  also been predicted to exist as solutions of NLSE in self-attracting BEC in the presence of periodic optical lattices~\cite{Yan03,Bai03}. Other solutions with discrete rotational symmetry and a complicated phase structure have been introduced in periodically modulated potentials, both in the framework of BEC and nonlinear optics~\cite{Cra02,Sak05,Ale07a, Wan08,Ost04a,Ost04b,Ale07b}. Some of these solutions are characterised by the presence of more than one phase singularity~\cite{Cra02,Sak05,Ale07a, Wan08}. Finally, other quasi-stationary solutions showing discrete symmetry in a homogeneous medium, known as necklace beams or soliton clusters have been introduced, and some of them show a non-trivial phase structure~\cite{Sol98,Sol00,Sol01,Des01,Des02,Des04}. Stationary solutions of this kind have been also obtained in an inhomogeneous media such as a photonic lattice~\cite{Kar04a,Kar04b,Yan05}.

In this paper, our aim is to study DV's with different possible configurations of singularities. We will analitically  obtain the behavior of a DV near the symmetry axis to show that they always present a singularity in the rotational axis which can be completely characterised by discrete group theory arguments. Next, we will show that they can present more than one single singularity. The positions of these singularities are related according to rules arising from discrete group theory arguments. We will also provide two numerical examples of DV's with more than one singularity to illustrate our results. 

\section{Theory}

To start with, let us introduce first some common definitions which we will be using in the text. If  $\psi(\mathbf{x})$ is a complex scalar solution of a wave equation defined in a two-dimensional domain, $\mathbf{x}\in\mathbb{R}^2$, then the winding number $\gamma$ of $\psi$
along a closed curve $\Gamma$ is given by the  contour
integral $
\gamma=\frac{1}{2\pi}\oint_{\Gamma}\nabla\phi\cdot d\mathbf{l}$, where $\phi$ is the phase of the complex field
    $\psi=|\psi|e^{i\phi}$. Let $\mathbf{x}_{0}$
be the position of a phase singularity of $\psi$. The topological charge of the phase singularity located at $\mathbf{x}_{0}$
is the winding number of the complex field $\psi$ for the smallest
closed curve containing $\mathbf{x}_{0}$.
That is, if $\Gamma_{\epsilon}$ is a family of closed curves containing
$\mathbf{x}_{0}$ parametrized by $\epsilon$ such that
$\lim_{\epsilon\rightarrow0}\Gamma_{\epsilon}=\mathbf{x}_{0}$ 
then $v\equiv\lim_{\epsilon\rightarrow0}\frac{1}{2\pi}\oint_{\Gamma_{\epsilon}}\nabla\phi\cdot
d\mathbf{l}$. Additionally, one can define another quantity, the total angular momentum, as $<\psi|\mathcal{L}_z|\psi^*>/<\psi|\psi^*>$, where $\mathcal{L}_z=(\vec{r}\times\vec{\bigtriangledown})_z$.

Vortices with circular symmetry can be written as $\psi(r,\theta )=g(r)e^{il\theta}$, where $ (r,\theta) $ are the polar coordinates.  They present well defined angular momentum since $\mathcal{R}_{\alpha}\psi=e^{il\alpha}\psi$, where $\mathcal{R}_{\alpha}=e^{i\mathcal{L}_z\alpha}$ is a continuous rotation of angle  $\alpha\in\mathbb{R}$ and  $\mathcal{L}_z=i\frac{\partial}{\partial \theta}$ is the generator of $\mathcal{O}(2)$ rotation group. Therefore,  vortices are eigenfunctions of the angular-momentum operator, satisfying $\mathcal{L}_z\psi=l\psi$. Note that the winding number and the topological charge of a singularity are phase-related concepts while the angular momentum is a symmetry-related concept. These circularly symmetric vortices present a single singularity at the origin and fulfill that $l=v=\gamma$, provided that $\gamma$ is calculated in a closed curve that surrounds the phase singularity. Also, they satisfy that $g(r)\sim a r^{|l|}+O\left(r^{|l|+2}\right)$ when $r\rightarrow0$ \cite{DesyatnikovPO2005}. 

For a DV, the winding number  $\gamma$ and the topological charge $v$, i.e. the phase-related concepts, are also well defined.  However angular momentum is not.  
Nevertheless, another symmetry-related concept, the angular pseudo-momentum $m$, have been defined for discrete symmetry media \cite{Fer05PRE}. It has been also demonstrated that this quantity is conserved during propagation \cite{Fer05PRE}. By construction, the angular pseudo-momentum $m$ completely defines
the representation of the $\mathcal{C}_{n}$ group to which $\psi$
belongs. It has been also shown that $m=v=\gamma$ for a DV with a single singularity \cite{Fer05PRL}. Nevertheless, to our knowledge, the relationship among these this three quantities has not yet been  established for all types of DV's. This relationship will permit us to study the number of phase singularities associated to a DV.

Let us show next  how the behavior of a DV near the symmetry axis can  be analitically determined in terms of the angular pseudomomentum $m$. This will permit us to establish that there is always a phase singularity of charge $m$ located at the symmetry axis.  To do this, let the function $\psi(r,\theta)$ be a stationary solution of the
2D nonlinear eigenvalue equation:
\beq
\left[L_{0}+L_{NL}(|\psi|)\right]\psi=\mu\psi
\label{eq:stationary_eq}
\eeq
 where $L_{0}=-\nabla_{t}^{2}+V(\mathbf{x})$ and $L_{NL}(|\psi|)$
be invariant operators
under the $\mathcal{C}_{n}$ point-symmetry group formed by discrete
rotations of order $n$ around the rotation axis. It has been
proved in references \cite{Fer05PRE,Fer05OE}
that if $\psi$ is a self-consistent solution of (\ref{eq:stationary_eq})
satisfying the symmetry condition
$|\psi(r,\theta+2\pi/n)|^{2}=|\psi(r,\theta)|^{2}$,
it can be expressed
as $\psi_{m,\alpha}(r,\theta)=e^{im\theta}u_{m,\alpha}(r,\theta)$,
where $u_{m,\alpha}(r,\theta)=u_{m,\alpha}(r,\theta+\frac{2\pi}{n})$,
$m$ is the angular pseudo-momentum and $\alpha$ is a band index.
The function $\psi$ is said to be a \emph{symmetric stationary solution}
of the 2D nonlinear eigenvalue equation \eqref{eq:stationary_eq}
if it satisfies the condition
$|\psi(r,\theta+2\pi/n)|^{2}=|\psi(r,\theta)|^{2}$. 

It has been also demontrated that the angular pseudo-momentum presents a cut-off related
with the order $n$ of discrete rotational symmetry of the medium. Particularly, it has been shown that
$|m|\le\frac{n}{2}$ for even $n$ and $|m|\le\frac{n-1}{2}$ for odd $n$ \cite{Fer05PRE,Fer05PRL}. Under these conditions it can be proved that the solutions of Eq. (\ref{eq:stationary_eq}) with $|m|=1,\dots,\frac{n}{2}-1$ for even $n$ and $|m|=1,\dots,\le\frac{n-1}{2}$ for odd $n$ cannot be  real. It is easy to see that the operator $L=\left[L_{0}+L_{NL}(|\psi|)\right]$ is Hermitian. Therefore its eigenvalues are real numbers. Since $L$ satisfies  $L=L^*$, solutions with $m=0$ or $m=\frac{n}{2}$ for even $n$ are real (up to a global phase), since they belong to one-dimensional irreducible representations. On the other hand, solutions with angular pseudo-momentum different from $m=0$ or $m=\frac{n}{2}$ for even $n$ correspond to complex solutions belonging to two-dimensional representations of the $\mathcal{C}_n$ group. Therefore, DV's are characterized by $|m|=1,\dots,\frac{n}{2}-1$ for even $n$ and $|m|=1,\dots,\frac{n-1}{2}$ for odd $n$. 

We leave the study of the solutions  with   $m=0$ or $m=\frac{n}{2}$ for future research.  For the rest, i.e., for DV's, we will obtain next the mathematical behavior near the symmetry axis.  Particularly, we will show that if the function $\psi$ is a symmetric stationary
solution of \eqref{eq:stationary_eq} and satisfies the following mathematical conditions
\begin{enumerate}
\item $|\psi(r,\theta)|\stackrel{r\rightarrow0}{\approx}|\psi_{0}|+\delta\psi(r,\theta)$, 
where $\psi_{0}\in\mathbb{R}$ and $\delta\psi\ll\psi_{0}$,
\item $L_{NL}(|\psi|)\stackrel{r\rightarrow0}{\approx}L_{NL}(|\psi_{0}|)+\delta
L_{NL}$, 
where $L_{NL}(|\psi_{0}|)\in\mathbb{R}$ and $\delta L_{NL}\ll
L_{NL}(|\psi_{0}|)$ 
, and 
\item $V(r,\theta)\stackrel{r\rightarrow0}{\approx}V_{0}+\delta V(r,\theta)$,
where $V_{0}\in\mathbb{R}$ and $\delta V\ll V_{0}$,
\end{enumerate}
then 
\[
\psi(r,\theta)\stackrel{r\rightarrow0}{\approx}ar^{m}e^{im\theta}
+\mathcal{O}(r^{m+1}),\]   
where $a\in\mathbb{R}$.  

From a physical point of view, these mathematical conditions establish that the amplitude, the nonlinear operator, and the potential $V$ have a smooth non-singular behavior near the symmetry axis. Therefore, the previous conditions are easily satisfied by DV's in discrete symmetry media such as the systems mentioned above. We will prove next that the mathematical behavior of DV's near the symmetry axis depends only on the angular pseudo-momentum. 

To demonstrate the above behavior of the function $\psi(r,\theta)$, we need to express it as 
$\psi_{m,\alpha}(r,\theta)=e^{im\theta}u_{m,\alpha}(r,\theta)$. 
The functions $u_{m,\alpha}(r,\theta)$ satisfy the following differential equation
\beqa
 && \left[-\frac{\partial^{2}}{\partial
     r^{2}}-\frac{1}{r}\frac{\partial}{\partial
     r}+\frac{m^{2}}{r^{2}}-i\frac{2m}{r^{2}}\frac{\partial}{\partial\theta}
-\frac{1}{r^{2}}\frac{\partial^{2}}{\partial\theta^{2}}\right]u_{m,\alpha}
\nonumber\\   
 &&
  +\left[V(r,\theta)+L_{NL}(|\psi|)\right]u_{m,\alpha}
=\mu_{m,\alpha}u_{m,\alpha}.  
\eeqa
Because of the periodic behavior of the wave function $u_{m,\alpha}(r,\theta)$
and the potential $V(r,\theta)$ we can expand them in Fourier
series in the angular variable as
\[
u_{m,\alpha}(r,\theta)=\sum_{k}e^{ikn\theta}u_{m,\alpha}^{k}(r),\,\,\,\,
V(r,\theta)=\sum_{k'}e^{ik'n\theta}V_{k'}(r),\]
and after performing the angular integrals we get the following set of
differential equations for the angular Fourier components $u_{m,\alpha}^{\bar{k}}(r)$
\beqa
&&\left[-\frac{d^{2}}{dr^{2}}-\frac{1}{r}\frac{d}{dr}
+\frac{(m+\bar{k}n)^{2}}{r^{2}}\right]u_{m,\alpha}^{\bar{k}}
 +  \sum_{k}V_{\bar{k}-k}(r)u_{m,\alpha}^{k}\nonumber \\ 
&&+\sum_{k}L_{NL}^{k-{\bar{k}}}(|\psi|)u_{m,\alpha}^{\bar{k}}
=\mu_{m,\alpha}u_{m,\alpha}^{\bar{k}}.
\label{eq:set_diff_eqs}
\eeqa
Assumptions (1), (2) and (3) for the limit $r\rightarrow0$,
 allow us to write $V(r,\theta)\approx V_{0}$,
$|\psi(r,\theta)|\approx\psi_{0}$, 
and $L_{NL}(|\psi|)\approx L_{NL}(|\psi_{0}|)$, where
$V_{0},\,\psi_{0},\,\mbox{and}\, L_{NL}(|\psi_{0}|)$ are real numbers. Therefore, in this limit,  
we obtain 
\beqas
V_{\bar{k}-k}(r)&=&\frac{1}{2\pi}\int_{0}^{2\pi}d\theta
e^{i(\bar{k}-k)n\theta}V(r,\theta)\\
&\stackrel{r\rightarrow0}{\approx}&
\delta_{\bar{k},k}V_{0}+\delta V_{\bar{k}-k}(r), 
\eeqas
and
\beqas
L_{NL}^{k-{\bar{k}}}(|\psi|)&=&\frac{1}{2\pi}\int_{0}^{2\pi}d\theta
e^{-i(\bar{k}-k)n\theta}L_{NL}(|\psi|)\\
&\stackrel{r\rightarrow0}{\approx}&\delta_{\bar{k},k}L_{NL}(|\psi_{0}|)+\delta 
L_{NL}^{k-\bar{k}}.
\eeqas

 Then, the set of differential equations \eqref{eq:set_diff_eqs}  for the angular Fourier components  behave
for $r\rightarrow0$ as 
\beqa
&&\left[-\frac{d^{2}}{dr^{2}}-\frac{1}{r}\frac{d}{dr}
+\frac{(m+\bar{k}n)^{2}}{r^{2}}\right]u_{m,\alpha}^{\bar{k}} \nonumber\\ 
&& \stackrel{r\rightarrow0}{\approx}  
\left(\mu_{m,\alpha}-V_{0}-L_{NL}(|\psi_{0}|)\right)u_{m,\alpha}^{\bar{k}}
= K_{m,\alpha}u_{m,\alpha}^{\bar{k}}
\eeqa
%
which is the same equation (but in the linear case) obtained for circularly symmetric vortices of the form $\psi(r,\theta )=u(r)e^{il\theta }$ carrying angular momentum $l=m+\bar{k}n$.   As discussed in \cite{DesyatnikovPO2005}, the behavior of a solution of this type close to the origin is given by $u(r)\stackrel{r\rightarrow0}{\approx}Ar^{|l|}+O(r^{|l|+1})$. Therefore,  the solution of the previous equation behaves in the vortex core region  as: 

\beqa
u_{m,\alpha}^{\bar{k}}(r)\stackrel{r\rightarrow0}
{\approx}ar^{\left|m+\bar{k}n\right|}+O(r^{\left|m+\bar{k}n\right|+1}).   
\label{limitu}
\eeqa


We now analized which are the dominant terms in the limit $r\rightarrow 0$ in the Fourier expansion of $\psi_{m,\alpha}(r,\theta)$.
\beqa
\psi_{m,\alpha}(r,\theta)=e^{im\theta}[\sum_k e^{ikn\theta}u^k_{m,\alpha}(r)]
\label{expansionpsi}
\eeqa
In the limit $r\rightarrow 0$ the dominant term in the expression above will be given by that which minimizes the exponent of the leading term in the expansion (\ref{limitu}.)
The behavior of $\eta(k)=|m+kn|$ has to be then analysed for
different $m$ and $k$ values in order to find the dominant contribution, i.e., that given by the lowest value of $\eta(k)$.

Let us first consider  that, according to the cut-off theorem, $|m|\le \frac{n}{2}$ for $n$ even  and $|m|\le \frac{n-1}{2}$ for  $n$ odd.  Besides, as stated above, for DV's, $m\neq0$ and $m \neq \frac{n}{2}$, and then 
$|m|=1,\cdots,\frac{n}{2}-1$ for $n$  even, or $|m|=1,\cdots,\frac{n-1}{2}$ for odd $n$. Therefore, it is easy to see that $|kn|>|m|$, for $|k|=1,2,\cdots$ and, of course, $|kn|<|m|$ for $k=0$. With this in mind, let us see that $\eta(0)<\eta(k)$ for all possible values of $m$ and $k$. 

Let us first  consider that $m>0$ and $k>0$. In this case, we have $\eta(k)=|m+kn|=m+kn$, for $k=0,1,\dots$, since both quantities are positive.  Hence, this gives $\eta(0)<\eta(k)$, for $k=1,2,\dots$. Next, let us consider that $m>0$ and $k<0$. For $|k|=1,2,\cdots$, we have $\eta(k)=|m+kn|=|k|n-m$ since $|kn|>|m|$. Then, this gives $\eta(k)<\eta(k+1)$. On the other hand, for $k=0$, we have $\eta(0)=m$. Finally,  let us show that $m<|k|n-m$. For DV's it is always satisfied that $|m|<\frac{n}{2}$ in all cases. Then, we obtain $2m<|k|n$ for $k\neq0$ and consequently  $m<|k|n-m$. Then,  this gives $\eta(0)<\eta(k)$. 

Now, let us consider that $m<0$ and $k\geq0$. Then, we have $\eta(0)=|m|$ and    $\eta(k)=|m+kn|=kn-|m|$, for $k=1,2,\cdots$ since $kn>|m|$ and then  $kn-|m|$ is always a positive value. Then,  this gives $\eta(k)<\eta(k+1)$ for $k=1,2,\cdots$. On the other hand, since for DV we have $2|m|<kn$ for $k=1,2,\cdots$, we obtain $|m|<kn-|m|$. Therefore, this gives $\eta(0)<\eta(k)$ for $k=1,2,\cdots$. Finally, if $m<0$ and $k\leq0$ then we have $\eta(0)=|m|$ and   $\eta(k)=|m+kn|=|kn|+|m|$, for $k=1,2,\cdots$ since both quantities are negative.  Then, obviously $\eta(0)<\eta(k)$, is satisfied for $|k|=1,2,\cdots$.

In conclusion, for all the possible values of $m$ that a DV can present, the dominant contribution to the wave
function arises from the term given by $k=0$ and, consequently, and according to Eq.~(\ref{limitu}) and (\ref{expansionpsi}) the wave function behaves like \[
\psi_{m,\alpha}(r,\theta)\propto e^{im\theta}r^{|m|}\]
 in the limit $r\rightarrow0$.

Therefore, it can be easily proved that there exists always a phase singularity of charge $m$ located at
the point
$\mathbf{x}_{r}$ where the rotation axis intersects the 2D plane, i.e. the topological charge of this singularity is $
v=m$. 

Once this has been established, let us go into the relationship between $m$, the winding number, and the topological charge of a singularity in depth. The winding number of the symmetric stationary solution
$\psi$ can be calculated using any closed curve. 
Let us consider a closed curve $\Gamma$ that surrounds the point $\mathbf{x}_{r}$
and let us assume that the winding number of the symmetric stationary solution
$\psi$ is such that $\gamma\neq m$. Then:

1. There exists always a phase singularity of charge $m$ located
\emph{on-axis,} and

2. There must exist a number of \emph{off-axis} singularities inside the
closed curve $\Gamma$ fulfilling the condition:

\beqa
\gamma=m+\sum_{j=1}^{V}v_{j},
\label{eq:rel_gamma_vj_m}
\eeqa
 where $v_{j}$ is the topological charge of the $j$th phase singularity and $V$ is the total number of singularities. 

3. Additionally, there could exist a number of off-axis phase singularity pairs with opposite
charges (vortex-antivortex pairs) inside the closed curve $\Gamma$. 

This result is obtained after spliting the path integral in the definition of $\gamma$, in $V+1$ path integrals, all of them related with the topological charge of each singularity. The integral around the rotational axis will offer topological charge equal to $m$. The rest must be related with the existence of $V$ off-axis singularities or vortex-antivortex pairs that contribute with a null net charge to the integral.

Alternatively, let us consider a closed curve $\Gamma$ that surrounds the point 
 $\mathbf{x}_{r}$  for which the winding number of the symmetric stationary solution
$\psi$ is such that $\gamma=m$. Then, using similar arguments, either:

1. There exists \emph{only} one phase singularity of charge $m$ located
on-axis, or,

2. There exist one phase singularity of charge $m$ located on-axis
and a number of off-axis phase singularity pairs with opposite
charges (vortex-antivortex pairs) inside the closed curve $\Gamma$. 

Hence, we have proved that it can be stated that a DV presents a phase singularity of topological charge $m$ located at the rotational axis and a number of off-axis phase singularities verifying Eq.~\ref{eq:rel_gamma_vj_m}.

Let us show now that  
the position of these off-axis phase singularities of $\psi$ is always symmetric with respect
to the rotation axis. Let us assume that a phase singularity is located at a point $\mathbf{x}_0$ different from the point $\mathbf{x}_r$, where the symmetry axis intersects the transverse plane. Then, the modulus of the function at this point vanishes, $|\psi(\mathbf{x}_0)|=0$ and the argument is not defined. Besides,  as stated above, the function $\psi$ of a DV must belong to one of the represetations of the group~\cite{Fer04,Fer05PRE}.  Then, if one transforms $\psi$ according to the rotational elements of the group, i.e., the transformations $C_{n}^t=e^{it\frac{2\pi}{n}}$, $t=1,\dots,n-1$, then the transformed function must be:
\begin{equation}
\label{eq:transform}
\psi_t=e^{itm\frac{2\pi}{n} }\psi.
\end{equation}
Then, the transformed function and the function itself differs only in a fixed phase, i.e. $|\psi_t|=|\psi|$ and $\angle{\psi_t}=\angle{\psi}+\alpha_t$, where $\alpha_t=tm\frac{2\pi}{n}$.  

Let us apply the rotational elements of the group $C_{n}^t$, $t=1,\dots,n-1$ to the function at the point $\mathbf{x}_0$. The transformed function is the function $\psi$  evaluated at a set of  rotated points $\mathbf{x}_t$ given by  $(r_t,\theta_t)=(r_0,\theta_0+t\frac{2\pi}{n})$, $t=1,\dots,n-1$, where we have used polar coordinates and where $(r_0,\theta_0)$ are the polar coordinates of the point $\mathbf{x}_0$. According to Eq.~(\ref{eq:transform}), if $|\psi(\mathbf{x}_0)|=0$  then $|\psi_t|=0$ for $t=1,\dots,n-1$. Consequently, the modulus of the function vanishes in this set of points. Let us show that there is a phase singularity located in each of these points.

To do so, let us consider  points $\mathbf{x}_0^{\theta}$ such that $\mathbf{x}_0^{\theta}=\mathbf{x}_0+\mathbf{x}_{(\epsilon,\theta)}$, where $|\mathbf{x}_{(\epsilon,\theta)}|=\epsilon $ with $\epsilon$ an infinitesimally  small quantity and $ \angle{\mathbf{x}_{(\epsilon,\theta)}}=\theta\in[0,2\pi]$. Taking into account the properties of a phase singularity, the phase of the function is increased or decreased in an integer multiple of $2\pi$ along the circle $\xi_0$ obtained after fixing $\epsilon$ and varying $\theta$ from 0 to $2\pi$. Now, let us apply the rotational elements of the group  $C_{n}^t=e^{it\frac{2\pi}{n}}$, $t=1,\dots,n-1$ to the function in all the points  in a circle such as $\xi_0$. The transformed functions are the function $\psi$ evaluated at  points $\mathbf{x}_t^{\theta}= \mathbf{x}_t+\mathbf{x}_{(\epsilon,\theta_t)}$ where $\theta_t=\theta+t\frac{2\pi}{n}$ of circles $\xi_t$.  Taking into account Eq.~(\ref{eq:transform}), the phase of the transformed function at each point of the circle $\xi_t$ and of the function itself at each point of the circle $\xi_0$ differs only in a fixed number $\alpha_t$, which is the same for all the points at $\xi_t$. Consequently, the phase of the function still increases or decreases in an integer multiple of $2\pi$ along the circles $\xi_t$ for any $\epsilon$. 

Then, in the one hand, the function vanishes at the points $\mathbf{x}_t$, and, in the other hand, the phase of the function is increased or dicreased an integer multiple of $2\pi$ along any circle surrounding these points. This proves that, if there exist an  off-axis phase singularity, there are other $n-1$ \emph{off-axis} phase singularities distributed symmetrically with respect to the rotation axis. 

Then, one can re-write Eq.~(\ref{eq:rel_gamma_vj_m}) as:

\beqa
\gamma=m+\sum_{k=1}^{K}n\,v_{k},
\label{eq:rel_gamma_vk_m}
\eeqa
 where $v_{k}$ is the topological charge of each of the $n$ phase singularities related by the symmetry conditions, and $K$ is the total number of these \emph{rings} of singularities.

\section{Numerical Results}

\begin{figure}
\begin{tabular}{cc}
 (a)&
\hspace{1cm}
(b)\tabularnewline
\includegraphics[%
  scale=0.65]{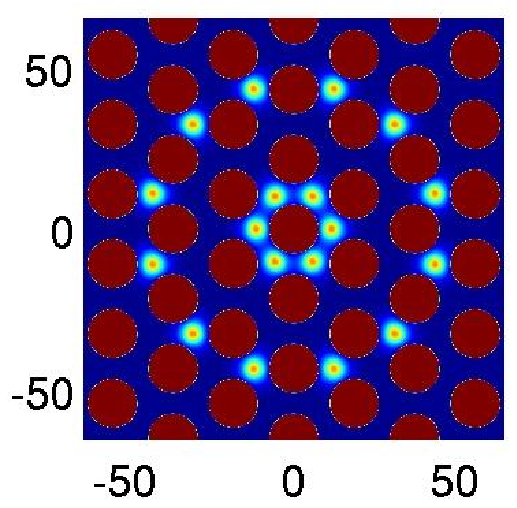} &
\includegraphics[%
  scale=0.65]{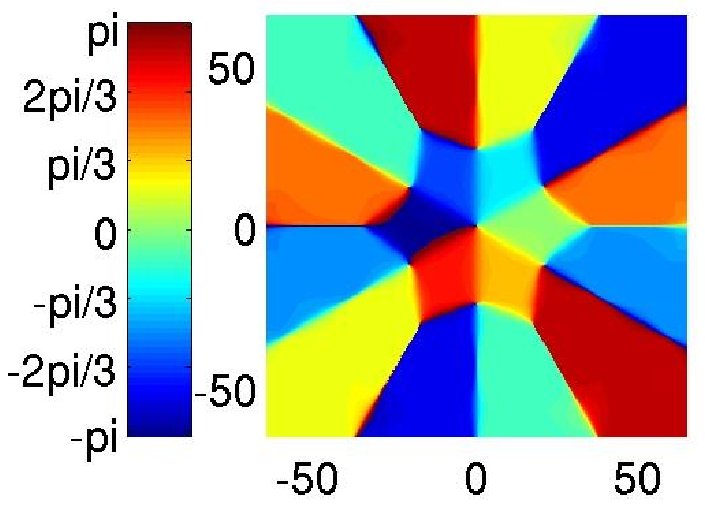} \tabularnewline
\end{tabular}
\caption{Amplitude (a) and phase (b) of a DV with angular pseudo-momentum $m=-1$ and more than one singularity. Red circles in (a) represent the air holes in the photonic crystal fiber. \label{fig:fig1}}

\end{figure}

\begin{figure}
\begin{tabular}{cc}
(a)&
\hspace{1cm}
(b)\tabularnewline
\includegraphics[%
  scale=0.65]{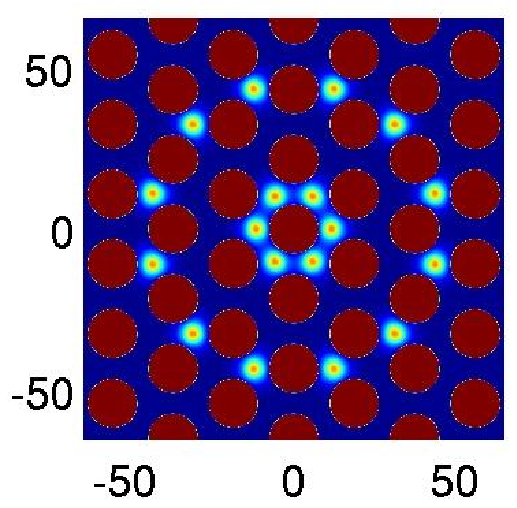} &
\includegraphics[%
  scale=0.65]{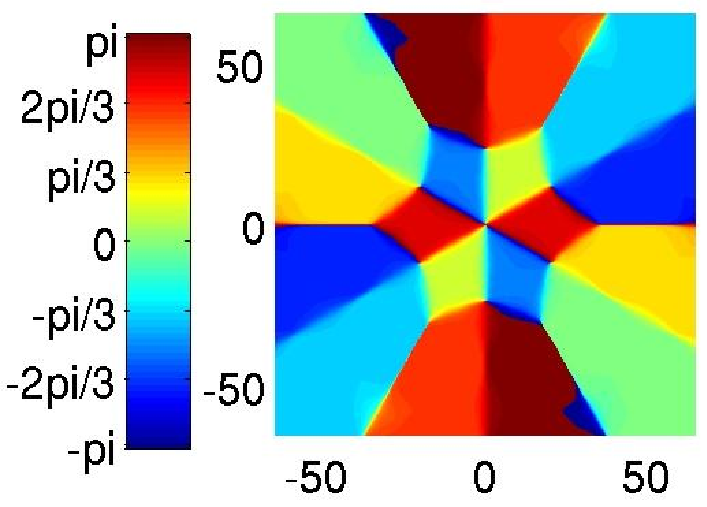} \tabularnewline
\end{tabular}
\caption{Amplitude (a) and phase (b) of a DV with angular pseudo-momentum  $m=-2$ and more than one singularity. Red circles in (a) represent the air holes in the photonic crystal fiber.  \label{fig:fig2}}

\end{figure}

Let us illustrate the previous results with numerical examples of optical DV with more than one singularity in photonic crystal fibers showing  $\mathcal{C}_{6v}$ discrete symmetry. In Ref.~\cite{Fer04,Fer05PRE}, DV's with just one singularity  in this kind of systems have been introduced. Additionally, the cut-off theorem for discrete symmetry media introduced in Ref.~\cite{Fer05PRL} states that  the only allowed values for the angular pseudo-momentum $m$ for a system with discrete symmetry of order $n=6$, are $m=\pm1$ or $m=\pm2$.  In Figs.~(\ref{fig:fig1}) and~(\ref{fig:fig2}) we present two stationary solutions of Eq.~(\ref{eq:stationary_eq}) with self-focusing non-linearity $L_{NL}(|\psi|)=|\psi|^2$. According with the transformations of the $\mathcal{C}_{6v}$ discrete symmetry group, these solutions show angular pseudo-momentum  $m=-1$ and $m=-2$, respectively. The numerical calculation of the winding number along a closed curve near the boundary of the numerical  solutions is $\gamma=5$ and $\gamma=4$, respectively. Then, in accordance with Eq.~(\ref{eq:rel_gamma_vj_m}) there are  $V=6$ off-axis singularities in each case, and an undetermined number of vortex-antivortex pairs. According to Eq.~(\ref{eq:rel_gamma_vk_m}), there is at least one ring of off-axis singularities. A thorough analysis of the phase of both stationary solutions shows, first, that there is a singularity in the symmetry axis. The calculation of the topological
charge of this singularity is $v=-1$ and $v=-2$, for each case, and therefore equal to the corresponding angular pseudo-momentum. On the other hand, there are a limited number of points where the off-axis singularities can be located, i.e., those where the phase seem to be undetermined. The calculation of the topological charge around these points shows that the singularities are located in the points marked with white circles in figures~(\ref{fig:fig1}b) and~(\ref{fig:fig2}b) and present topological charge $v=+1$. Therefore, Eq.~(\ref{eq:rel_gamma_vj_m}) is fulfilled, since $\gamma=5=-1+6$ for DV with $m=-1$, and $\gamma=4=-2+6$ for DV with $m=-2$.  Finally, it is easy to check for both solutions that the positions of these off-axis singularities are related according to the transformations of the discrete group.  

In conlusion, we have obtained analitically the behavior of a DV near the symmetry axis. With this result, we have been able to establish a general relationship between angular pseudo-momentum, winding number and topological charge for DV. This rule permits to study DV with more than one singularity in any discrete symmetry media. Additionally, we have shown that the positions of the off-axis singularities are related according to symmetry arguments. Finally, the results have been illustrated with two numerical examples of high-charged discrete vortices in a system with $\mathcal{C}_{6v}$ discrete symmetry. 

{\bf Acknowledgements}

This work was supported by Ministerio de Educaci\'on y Ciencia of Spain (projects TIN2006-12890 and FIS2005-01189) and Generalitat Valenciana (contract APOSTD/2007/052 granted to MZG). SS. is thankful to Generalitat Valencia for a visiting fellowship to Universitat de Val\`{e}ncia, where part of this work was done.

\end{document}